\documentclass[sigconf]{acmart}
\AtBeginDocument{%
  }
\usepackage{microtype}
\usepackage{graphicx}
\usepackage{pifont}
\usepackage{subcaption}
\usepackage{listings}
\usepackage{booktabs}       
\usepackage{amsfonts}       
\usepackage{nicefrac}       
\usepackage{multirow}
\usepackage{algorithm}
\usepackage{algorithmic}
\usepackage{amsmath}
\usepackage{mathtools}
\usepackage{amsthm}
\usepackage{makecell}

\usepackage{amsmath,amsfonts,bm}

\def\eqref#1{equation~\ref{#1}}

\def\1{\bm{1}}

\def\rmE{{\mathbf{E}}}

\def\rmI{{\mathbf{I}}}

\def\vzero{{\bm{0}}}

\def\vc{{\bm{c}}}
\def\vd{{\bm{d}}}

\def\vx{{\bm{x}}}

\def\vz{{\bm{z}}}

\DeclareMathAlphabet{\mathsfit}{\encodingdefault}{\sfdefault}{m}{sl}
\SetMathAlphabet{\mathsfit}{bold}{\encodingdefault}{\sfdefault}{bx}{n}

\def\gN{{\mathcal{N}}}

\def\gP{{\mathcal{P}}}

\usepackage{xcolor}
\usepackage{enumitem}
\usepackage{ulem}
\copyrightyear{2025} 
\acmYear{2025} 
\setcopyright{acmlicensed}
\acmConference[MM '25]{Proceedings of the 33rd
ACM International Conference on Multimedia}{October 27--31, 2025}{Dublin,
Ireland}
\acmBooktitle{Proceedings of the 33rd ACM International Conference on
Multimedia (MM '25), October 27--31, 2025, Dublin, Ireland}
\acmDOI{10.1145/3746027.3755377}
\acmISBN{979-8-4007-2035-2/2025/10}

\begin{document}

\title{MEDIC: Zero-shot Music Editing with Disentangled Inversion Control}

\author{Huadai Liu}
\authornote{Equal contribution.}
\email{liuhuadai@zju.edu.cn}
\orcid{0009-0004-5782-5641}
\affiliation{%
    \institution{HKUST}
  \institution{Alibaba Group}
  \city{Hangzhou}
  \country{China}
}

\author{Jialei Wang}
\authornotemark[1]
\email{3220101016@zju.edu.cn}
\orcid{0009-0002-2786-3896}
\affiliation{%
  \institution{Zhejiang University}
  \city{Hangzhou}
  \country{China}
}

\author{Xiangtai Li}
\email{xiangtai94@gmail.com}
\orcid{0000-0002-0550-8247}
\affiliation{%
  \institution{ByteDance Inc.}
  \city{Singapore}
  \country{Singapore}
}

\author{Wen Wang}
\email{wwang.969803@gmail.com}
\orcid{0000-0002-0356-1968}
\affiliation{%
  \institution{Alibaba Group}
  \city{Sunnyvale}
  \state{California}
  \country{United States}
}

\author{Qian Chen}
\email{lukechan1231@gmail.com}
\orcid{0000-0001-6939-7438}
\affiliation{%
  \institution{Alibaba Group}
  \city{Hangzhou}
  \country{China}
}

\author{Rongjie Huang}
\email{rongjiehuang@zju.edu.cn}
\orcid{0000-0002-1695-9000}
\affiliation{%
  \institution{Zhejiang University}
  \city{Hangzhou}
  \country{China}
}

\author{Yang Liu}
\email{22160155@zju.edu.cn}
\orcid{0009-0005-8246-0741}
\affiliation{%
  \institution{Zhejiang University}
  \city{Hangzhou}
  \country{China}
}

\author{Jiayang Xu}
\email{jiayangxu@zju.edu.cn}
\orcid{0009-0002-2557-2438}
\affiliation{%
  \institution{Zhejiang University}
  \city{Hangzhou}
  \country{China}
}

\author{Zhou Zhao}
\email{zhaozhou@zju.edu.cn}
\orcid{0000-0001-6121-0384}
\affiliation{%
  \institution{Zhejiang University}
  \city{Hangzhou}
  \country{China}
}

\author{Wei Xue}
\email{weixue@ust.hk}
\authornote{Corresponding author.}
\orcid{0000-0002-4942-7748}
\affiliation{%
  \institution{HKUST}
  \city{Hong Kong}
  \country{China}
}

\renewcommand{\shortauthors}{Huadai Liu et al.}

\begin{abstract}
  Text-guided diffusion models revolutionize audio generation by adapting source audio to specific text prompts. However, existing zero-shot audio editing methods such as DDIM inversion accumulate errors across diffusion steps, reducing the effectiveness. Moreover, existing editing methods struggle with conducting complex non-rigid music edits while maintaining content integrity and high fidelity.
    To address these challenges, we propose MEDIC, a novel zero-shot music editing system based on innovative \textbf{Disentangled Inversion Control (DIC)} technique, which comprises \textbf{Harmonized Attention Control} and \textbf{Disentangled Inversion}. Disentangled Inversion disentangles the diffusion process into triple branches to rectify the deviated path of the source branch caused by DDIM inversion. Harmonized Attention Control unifies the mutual self-attention control and the cross-attention control with an intermediate Harmonic Branch to progressively generate the desired harmonic and melodic information in the target music. We also introduce \textbf{ZoME-Bench}, a comprehensive music editing benchmark with 1,100 samples covering ten distinct editing categories. ZoME-Bench facilitates both zero-shot and instruction-based music editing tasks. Our method outperforms state-of-the-art inversion techniques in editing fidelity and content preservation. The code and benchmark will be released. Audio samples are available at \url{https://melody-edit.github.io/}.
\end{abstract}

\ccsdesc[500]{Applied computing~Sound and music computing}
\ccsdesc[500]{Computing methodologies~Natural language generation}

\keywords{Zero-shot Music Editing, Inversion Techniques, Diffusion Models}

\maketitle

\begin{figure*}[htbp]
    \centering
    \begin{subcaptionblock}{0.45\textwidth}
        \centering
        \includegraphics[width=\linewidth]{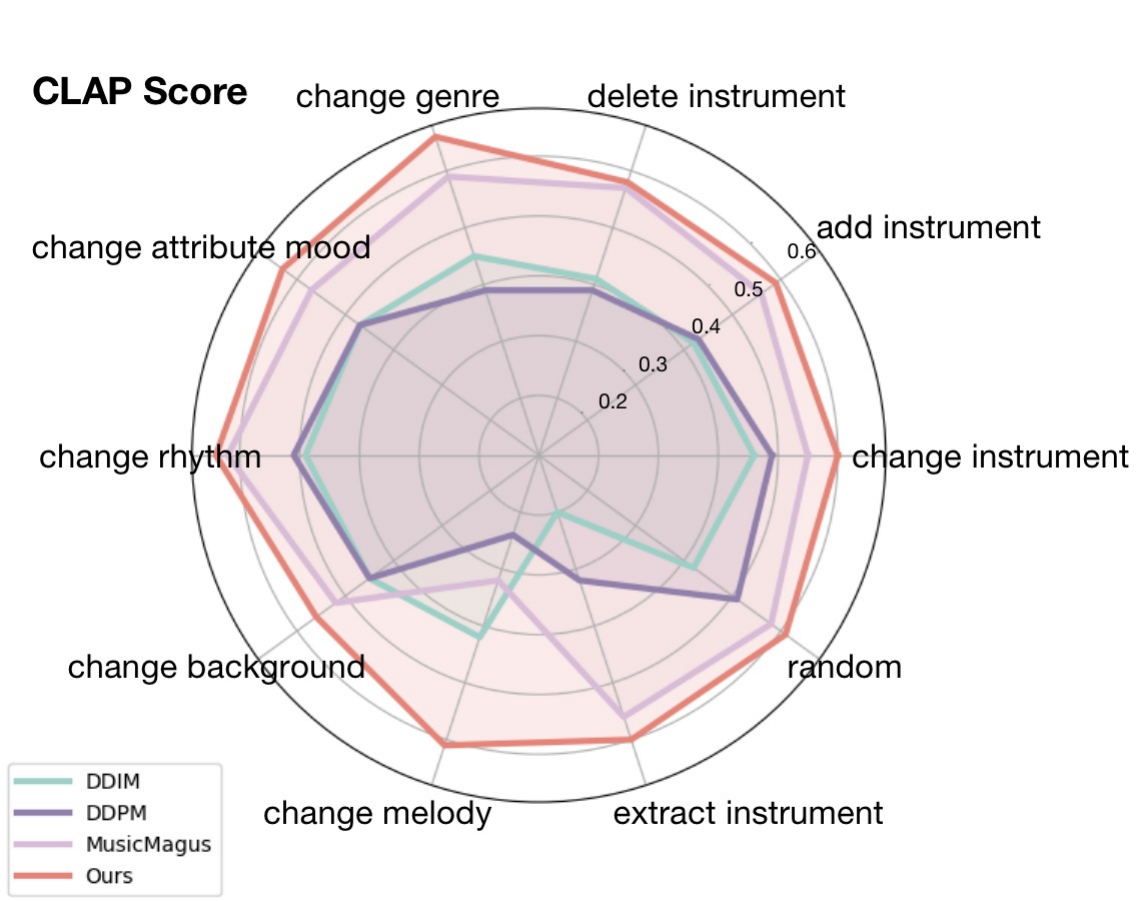}
        \caption{Spider chart for CLAP score comparisons.}
        \label{fig1-a}
    \end{subcaptionblock}
    \hfill
    \begin{subcaptionblock}{0.45\textwidth}
        \centering
        \includegraphics[width=\linewidth]{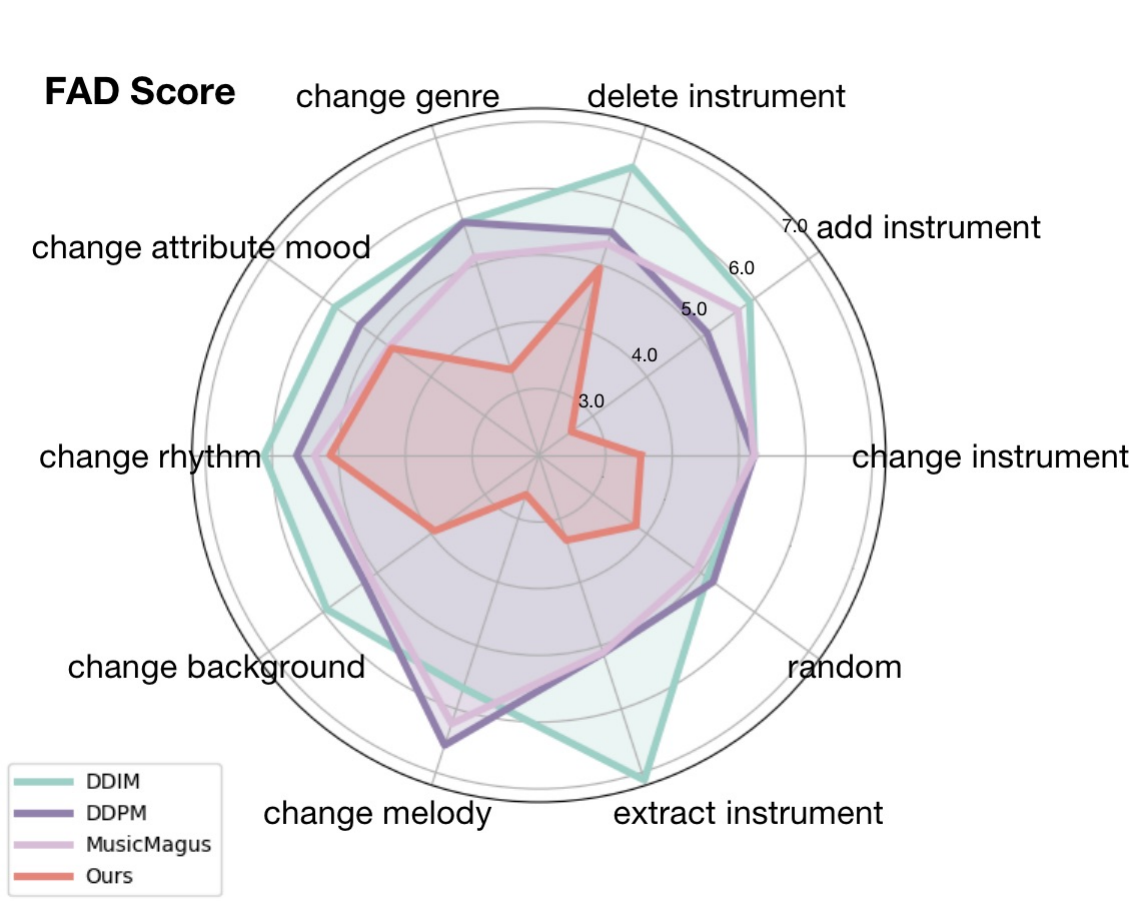}
        \caption{Spider chart for FAD score comparisons.}
        \label{fig1-b}
    \end{subcaptionblock}
    \vspace{-1mm}
    \caption{Comprehensive zero-shot music editing performance evaluation on the ZoME-bench. We present spider charts of CLAP scores (higher is better) corresponding to \textit{editing fidelity} and FAD scores (lower is better) corresponding to \textit{source-anchored attribute consistency} across 10 editing tasks (e.g., change genre) for DDPM-Friendly~\citep{manor2024zero}, DDIM Inversion~\citep{song2020denoising}, MusicMagus~\citep{zhang2024musicmagus}, and our MEDIC with Disentangled Inversion Control.}
    \label{fig:spider}
    \label{fig1}
\end{figure*}

\section{Introduction}
\label{1}



%
Text-guided diffusion models~\citep{song2020denoising,song2020improved,peebles2023scalable} have made great progress in audio generation~\citep{evans2024fast,evans2024long}, leveraging their impressive capability of generating realistic and varied outputs. 
These models~\citep{liu2023audioldm2,huang2023makeanaudio,liu2024audiolcm} provide the foundation for prompt-based music editing, offering new opportunities to modify audio landscapes for specific \textit{text prompts}. 
Early music editing strategies rely on training models from scratch~\citep{copet2023simple,agostinelli2023musiclm} or test-time optimization~\citep{paissan2023audio,plitsis2024investigating}, hence they are hampered by intensive computational demands. 
%
Recent works~\citep{manor2024zero,zhang2024musicmagus}
 have advanced zero-shot music editing through Denoising Diffusion Implicit Models (DDIM)~\citep{song2020denoising} and Denoising Diffusion Probabilistic Models (DDPM)~\citep{ho2020denoising} inversion techniques, but challenges remain. 



%
There are two objectives for music editing, that is, \textbf{edit fidelity} - ensuring the editing aligns with the provided instructions - and \textbf{essential content preservation} - maintaining specified musical properties from the source prompt while modifying only designated attributes. For instance, when transforming a melancholic piano composition into an electric guitar arrangement, edit fidelity demands accurate instrument substitution while essential content preservation ensures sustained emotional resonance through retained tempo and recurring minor chord sequences.
Balancing these two objectives poses a great challenge since it involves a careful exchange of information between the source and target branch in diffusion processes; however, existing inversion methods such as DDIM prove sub-optimal for conditional diffusion models to address this challenge~\citep{mokady2023null}. 
%
Enhanced versions of edit-friendly DDPM inversion~\citep{hubermanspiegelglas2024edit} make strides in content preservation by imprinting the source onto the noise space. However, this method comes at the expense of reduced modification capabilities due to noise reduction. 


In this work, we first examine the shortcomings of the DDIM inversion approach. 
Our comprehensive analysis indicates that while techniques such as DDIM inversion provide a foundation for audio editing, they lack precision and may compromise the integrity of the original audio. The primary issue stems from the assumption of \textit{perfect reversibility} in the ordinary differential equation (ODE) process, which is frequently violated during text-conditional editing. This issue leads to distortions during the inversion. Although the implementation of Classifier-free Guidance (CFG)~\citep{classifier-free} aims to improve text adherence, CFG inadvertently amplifies the accumulated errors from the inversion process.
%

\begin{figure*}[htbp]
    \centering
    \vspace{-1mm}
    \includegraphics[width=.99\textwidth]{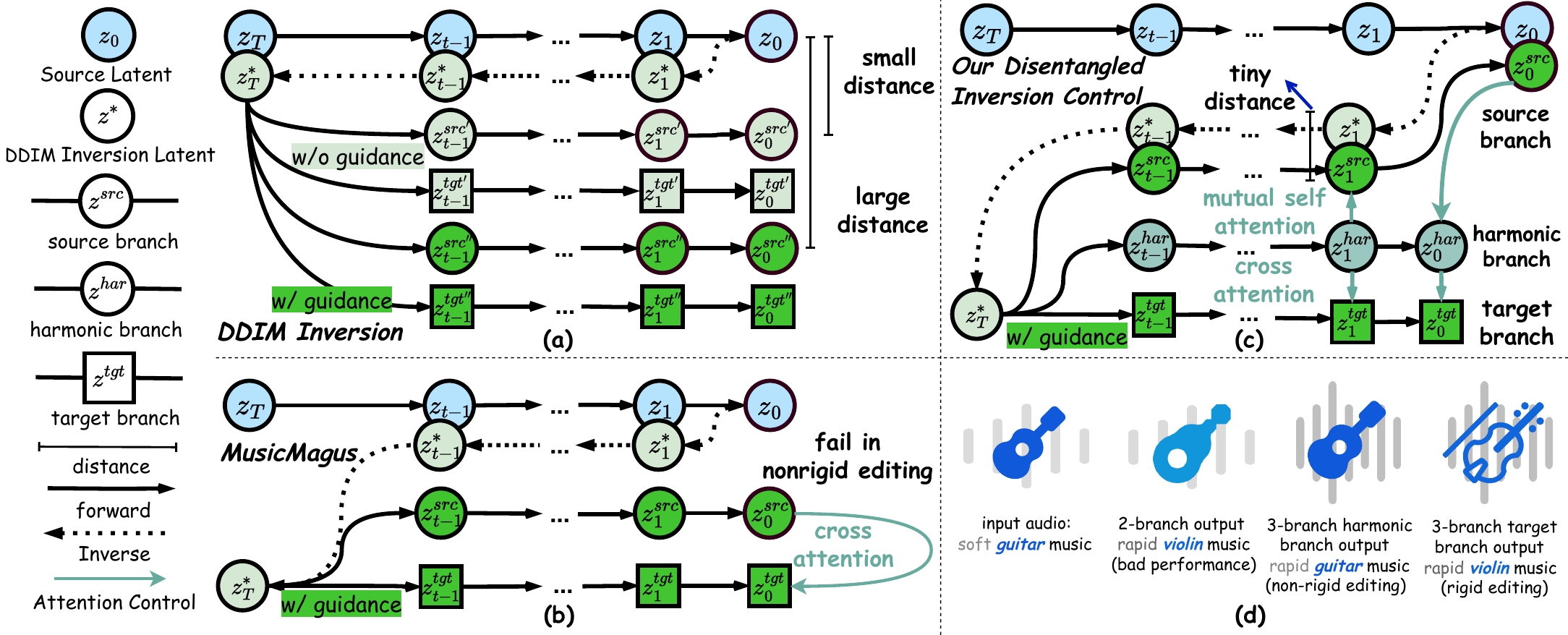}
    \vspace{-2mm}
    \caption{Comparisons between our method and major baselines based on two-branch inversion techniques, including DDIM Inversion~\citep{song2020denoising} and MusicMagus~\citep{zhang2024musicmagus}. (a) Framework of DDIM Inversion, showing configurations with and without classifier-free guidance. (b) Framework of MusicMagus, which incorporates cross-attention control. (c) Framework of our method, featuring Disentangled Inversion Control. (d) An illustration comparing the output of the two-branch techniques with the progressive output of our triple-branch method.}
    \vspace{-5mm}
    \label{fig:framework}
\end{figure*}

Recently, attention control~\citep{cao2023masactrl,hertz2022prompt} has shown promise in achieving high fidelity and essential content preservation. For instance, MusicMagus~\citep{zhang2024musicmagus} introduces Cross-Attention Control for fine-grained music manipulation of rigid tasks. Note that in this paper, \textbf{rigid music editing} refers to structural modifications requiring recomposition of core musical relationships, such as the instruments changes, genre transformation, macro-harmonic restructuring, etc; whereas, \textbf{non-rigid music editing} involves parametric adjustments preserving original musical relationships, including 
adjusting the beat, melody, pitch, rhythm, and other subtle aspects, which generally involves more microscopic editing.
Nevertheless, attention control methods fail to resolve the issues of accumulated errors and struggle to achieve accurate editing for both rigid and non-rigid tasks, as illustrated in Figure~\ref{fig:framework}.
%

In this work, to bridge this gap, we introduce a zero-shot music editing system \textbf{MEDIC} based on an innovative \textit{Disentangled Inversion Control} technique, which comprises two components of \textit{Harmonized Attention Control} and \textit{Disentangled Inversion}.
Cross-attention control~\citep{hertz2022prompt} and mutual self-attention control~\citep{cao2023masactrl} have demonstrated robust editing capabilities for rigid and non-rigid image editing tasks, respectively. However, simply combining these two approaches sequentially for music editing can result in sub-optimal performance, particularly in the original dual-branch setup, where it struggles with global attention refinement. To address this issue, we propose \textit{Harmonized Attention Control}, which unifies cross-attention control and mutual self-attention control by introducing an intermediate branch named \textit{Harmonic Branch}, designed to progressively modify both rigid and non-rigid attributes in music.
%
%
Furthermore, we disentangle the diffusion process into triple branches and correct the deviation path caused by CFG in the source branch, which affects the essential content preservation. The other branches remain unchanged to ensure high edit fidelity. 

 Due to the lack of standardized benchmarks in music editing, we also introduce a new benchmark \textbf{ZoME-Bench}, consisting of 1,100 audio clips in 10 rigorously curated editing categories across rigid and non-rigid tasks.
 Each entry is carefully assembled, comprising a source prompt, a source audio, a target text prompt, human instruction, and blended words intended for editing. We then extensively evaluate MEDIC and baselines on ZoME-Bench and other datasets.
 %
 %
Our contributions can be summarized as follows. 
\begin{itemize}[leftmargin=*,noitemsep]
\item We introduce a novel, \textbf{training-free} methodology called \textbf{Disentangled Inversion Control (DIC)}, designed to facilitate consistent manipulations of musical elements and intricate non-rigid editing tasks. We develop a zero-shot music editing system \textbf{MEDIC} based on DIC.
\item  Disentangled Inversion Control includes two critical algorithmic designs. (a) A \textbf{Harmonized Attention Control} framework is introduced to unify cross-attention and mutual self-attention control, which enables both rigid and non-rigid editing. (b) \textbf{Disentangled Inversion Technique} is proposed to achieve superior results with negligible inversion error by branch disentanglement and correction, aiding in accurately editing the music while preserving the content information. 
\item We introduce \textit{ZoME-Bench}, a new benchmark for music editing complete with comprehensive evaluation metrics. It consists of 1,100 audio clips categorized into 10 rigorously curated editing tasks, encompassing both rigid and non-rigid tasks.
\item Experimental results on ZoME-Bench indicate that MEDIC outperforms competitive baselines, achieving significant improvements in edit fidelity and essential content preservation, as depicted in Figure~\ref{fig:spider}. Moreover, MEDIC achieves state-of-the-art performance under the \textit{variable-length music editing settings} of the commonly used MusicDelta dataset.
\end{itemize}

\section{Related Work}

\noindent
\textbf{Text-based Audio editing.}
Some prior text-based audio editing studies utilize diffusion models to manipulate audio content according to the target prompt provided~\citep{paissan2023audio,plitsis2024investigating,han2023instructme}. 
The two primary challenges (i.e. editing fidelity and essential content preservation) command intense focus. 
%

Existing methodologies~\citep{novack2024ditto,wu2024music} for addressing these intricate challenges typically follow one of three paths. The first involves attempts to develop end-to-end editing models~\citep{copet2023simple,agostinelli2023musiclm,chen2024musicldm} that employ diffusion processes. 
However, these efforts are often hampered by indirect training strategies or a lack of comprehensive datasets. The second path involves test-time optimization strategies that utilize large pre-trained models for editing~\citep{paissan2023audio,plitsis2024investigating}. Despite their versatility, these methods are often burdened by the significant computational demands of fine-tuning diffusion models or optimizing text embeddings for signal reconstruction. Some methods choose to employ both strategies~\citep{kawar2023imagic}, which further increases the computational load. The third path involves inversion techniques, which typically use DDPM~\citep{hubermanspiegelglas2024edit,wu2023latent} or DDIM~\citep{song2020denoising,zhang2024musicmagus} inversion strategies to extract diffusion noise vectors that match the source signal.
Considering its rapid and intuitive zero-shot editing capabilities, in this work, we choose inversion techniques as our primary research framework. 
Different from existing inversion strategies, we propose a new inversion technique named Disentangled Inversion Control, which disentangles the diffusion process into triple branches with both mutual self-attention control and cross-attention control to achieve accurate editing while preserving structural information.


\noindent
\textbf{Inversion Techniques.}
The field of image inversion techniques has experienced significant progress in recent years~\citep{brooks2023instructpix2pix,kim2022diffusionclip,parmar2023zero,dhariwal2021diffusion}. 
Although DDIM inversion proves to be effective for unconditional diffusion models~\citep{song2020denoising,yang2025emovoice,liu2025speech}, its limitations become apparent when applied to text-guided diffusion models, particularly when classifier-free guidance is necessary for meaningful editing. 
Various solutions have been proposed to address these challenges~\citep{mokady2023null,tumanyan2022plugandplay}. 
For example, Negative-Prompt Inversion strategically assigns conditioned text embeddings to Null-Text embeddings, effectively reducing potential deviation during editing. 
In contrast, Edit-Friendly DDPM provides an alternative latent noise space via modified DDPM sample distributions, promoting the successful reconstruction of the desired image~\citep{hubermanspiegelglas2024edit}.
%
%
%
Optimization-based inversion methods using specific latent variables have recently gained popularity~\citep{ju2023direct,kawar2023imagic}. 
These methods are designed to minimize the accumulated errors that result from the inversion of DDIM. 
Techniques such as Null-Text Inversion~\citep{mokady2023null} are promising, but introduce complexity and instability into the optimization process. 
Different from these inversion techniques, we introduce a plug-and-play method called Disentangled Inversion Control to separate branches which enables each branch to unleash its maximum potential individually, achieving superior performance with considerably fewer computational resources.
%


\section{Preliminaries And Analyses}
\label{3}
This section introduces the foundational concepts of DDIM sampling and classifier-free guidance as applied to diffusion models for text-guided audio synthesis. We further analyze the challenges associated with these methods.

\subsection{Diffusion Models}
\label{3.1}
Text-guided diffusion models aim to map a random noise vector $\vz_t$ and textual condition $\vc$ to an output audio $\vz_0$, corresponding to the given conditioning prompt. We train a denoiser network $\epsilon_{\theta}(\vz_t, t, \vc)$ to predict the Gaussian noise $\epsilon \in \gN(0,\rmI)$ following this objective:
\begin{equation}
    \underset{\theta}{min} \rmE_{\vz_0,\epsilon\in\gN(0,\rmI),t\in Uniform(1, T)} ||\epsilon - \epsilon_\theta(\vz_t, t, \vc)||^2
    \label{eq:1}
\end{equation}
\noindent where noise is added to the sampled data $\vz_0$ according to diffusion time step $T$. During inference, given a noise vector $\vz_T$, the noise $i$ is gradually removed by sequentially predicting it using a pre-trained diffusion model for $T$ steps.
To generate audio from given $\vz_T$, we employ the deterministic DDIM sampling, where $\alpha$ is hyperparameter:
\begin{equation}
    \vz_{t-1} = \frac{\sqrt{\alpha_{t-1}}}{\sqrt{\alpha_t}}\vz_t + \sqrt{\alpha_{t-1}}(\sqrt{\frac{1}{\alpha_{t-1}}-1} - \sqrt{\frac{1}{\alpha_t} - 1})\epsilon_\theta(\vz_t, t, \vc)
    \label{eq:2}
\end{equation}

\subsection{DDIM Inversion}
\label{3.2}
While diffusion models have superior characteristics in the feature space that can support various downstream tasks, it is hard to apply them to audio in the absence of natural diffusion feature space for non-generated audio. Thus, a simple inversion technique known as DDIM inversion is commonly used for unconditional diffusion models, predicated on the presumption that the ODE process can be reversed in the limit of infinitesimally small steps:
\begin{equation}
    \vz^*_t = \frac{\sqrt{\alpha_t}}{\sqrt{\alpha_{t-1}}}\vz^*_{t-1} + \sqrt{\alpha_t}(\sqrt{\frac{1}{\alpha_t} - 1} - \sqrt{\frac{1}{\alpha_{t-1}} - 1})\epsilon_\theta(\vz^*_{t-1},t-1)
    \label{eq:3}
\end{equation}
\noindent where $\vz^*$ denotes DDIM inversion latent. However, in most text-based diffusion models, this presumption cannot be guaranteed, resulting in a perturbation from $\vz_t$ to $\vz^*_t$ in Equation~\ref{eq:2}, Equation~\ref{eq:3} and Figure~\ref{fig:framework}(a).
Consequently, an additional perturbation from $\vz^*_t$ to $\vz^{src}_t$ arises when sampling an audio from $\vz^*_T$ as shown in Figure~\ref{fig:framework}(a):
\begin{equation}
    \vz_t = \sqrt{\alpha_t}\vz_0 + \sqrt{1 - \alpha_t}\epsilon
    \label{eq:4}
\end{equation}

\subsection{Classifier-free Guidance}
\label{3.3}
Classifier-free Guidance (CFG)~\citep{classifier-free} is proposed to overcome the limitation of weak text adherence in text-conditioned models. The modified noise estimation in CFG can be expressed as:
\begin{equation}
    \hat{\epsilon}_\theta(\vz_t,t,\vc,\varnothing) = \omega\cdot\epsilon_\theta(\vz_t,t,\vc) + (1 - \omega)\cdot\epsilon_\theta(\vz_t,t,\varnothing)
    \label{eq:5}
\end{equation}
where $\varnothing$ is the embedding of a null text. A higher guidance scale $\omega$, which is intended to strengthen the model's fidelity to the text prompt, inadvertently magnifies the accumulated inversion error. 
CFG further leads to another perturbation from $\vz^{src'}_t$ to $\vz^{src''}_t$ due to the destruction of the DDIM process and error augmentation, as depicted in Figure~\ref{fig:framework}(a). This issue becomes problematic in editing scenarios where precise control over audio synthesis is desired. 

\section{Methodology}
\label{4}
\subsection{Task Definition}
Despite significant work in text-to-audio generation models~\citep{huang2023makeanaudio,liu2024audiolcm,liu2023audioldm2,liu2024flashaudio}, particularly with the emergence of Latent Diffusion Models (LDM)~\citep{liu2023audioldm,chen2024musicldm}, research on zero-shot music editing remains limited. Given the source music $\vx^{src}_0$ and its corresponding text prompt $\gP$, zero-shot music editing seeks to leverage the capabilities of text-to-music generation models to directly modify $\vx^{src}_0$ and $\gP$, and synthesize the desired music $\vx^{tgt}_0$, which is aligned with the target edited text prompt $\gP^*$. We compress source audio signal $\vx^{src}_0$ into latent $\vz^{src}_0$ for inversion.

\subsection{Disentangled Inversion Control}
\label{4.1}
 Preliminaries and Figure~\ref{fig:framework} reveal that while techniques like DDIM inversion offer an editable base, they fall short of precision, potentially compromising essential content preservation. The implementation of Classifier-free Guidance (CFG) further amplifies the accumulated errors. 

In the landscape of prompt-based editing~\citep{dong2023prompt,kim2022diffusionclip,feng2023trainingfree}, grasping linguistic subtleties and enabling more granular cross-modal interactions remains a formidable challenge. ~\citet{hertz2022prompt} acknowledges that in image editing, the fusion between text and visual modalities occurs within the parameterized noise prediction network $\epsilon_\theta$, which leads to the development of various \textit{attention control} techniques that guide the target denoiser network $\hat{\epsilon}_\theta$ in the image domain to better align with target prompts. However, similar control mechanisms for non-rigid music editing are noticeably limited.

Taking these insights forward, we introduce \textbf{Disentangled Inversion Control (DIC)}, a novel approach to achieve both rigid and non-rigid music editing. 
DIC strategically disentangles the diffusion process as \textbf{triple branches} (i.e. source branch, harmonic branch, and target branch), and allows each branch to optimize its functionality. 
At the same time, the DIC strategy leverages our proposed \textbf{harmonized attention control} to facilitate targeted editing, thus aligning with the dual objectives of preserving the original audio essence and ensuring edit fidelity. 
Next, we first introduce \textit{Harmonized Attention Control} in Section~\ref{4.2} and discuss \textit{Disentangled Inversion} in Section~\ref{4.3}.


\subsection{Harmonized Attention Control Framework}
\label{4.2}

\begin{figure}[t]
    \centering
    \vspace{-1mm}
    \includegraphics[width=0.50\textwidth]{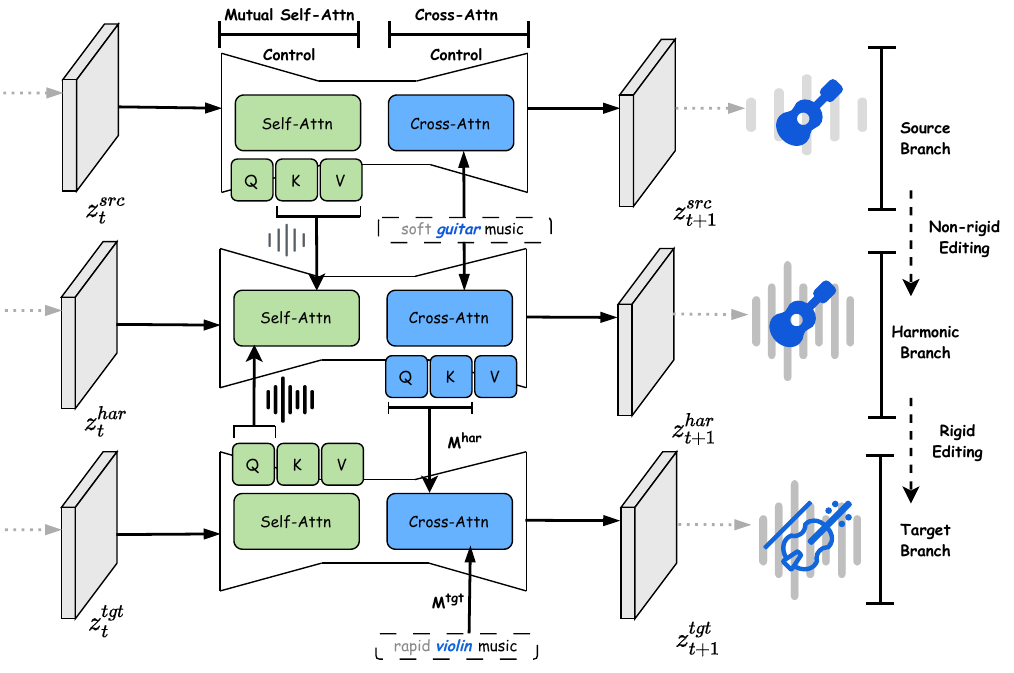}
    \vspace{-4mm}
    \caption{The Harmonized Attention Control (HAC) framework. HAC unifies cross-attention control and mutual self-attention control with an additional branch named \textit{Harmonic Branch} to host the desired composition and structural information in the target music.} 
    \vspace{-3mm}
    \label{fig:attention}
\end{figure}

The denoising architecture $\epsilon_\theta$ is structured as a sequence of fundamental blocks, each comprising a residual block~\citep{he2015deep} followed by self-attention and cross-attention modules~\citep{vaswani2023attention,dosovitskiy2020image,liu2023vit}. 
%
%
The proposed harmonized attention control (HAC) framework is depicted in Figure~\ref{fig:attention}.
We explore varying semantic transformations of audio content through harmonized attention control strategies - cross-attention control for rigid editings and mutual self-attention control for non-rigid editings. 
We introduce an intermediate Harmonic Branch to host the desired harmonic and melodic information in the target music. Below we elaborate on Cross-Attention Control, Mutual Self-Attention Control, and Harmonic Branch Integration within HAC.

\subsubsection{Cross-Attention Control}
\label{4.2.1}
Cross-attention Control (CAC) aims to inject the attention maps that are obtained from the generation with the original prompt $\gP$, into a second generation with the target prompt $\gP^*$. Motivated by Prompt-to-Prompt~\citep {hertz2022prompt}, We define CAC as \textit{Global Attention Refinement} and \textit{Local Attention Blend}.

\paragraph{\textbf{Global Attention Refinement}} At a given time step $t$, the attention map $M_t$ for both source and target branches is computed, averaging over all layers with respect to the noised latent $\mathbf{z}_t$. We employ an alignment function $A$ that maps each token index from the target prompt $\mathcal{P}^*$ to its equivalent in $\mathcal{P}$, or to None for unaligned tokens. The refinement action is:
\begin{equation}
    \text{Refine}(M^{src}_t,M^{tgt}_t, t) = 
    \begin{cases}
        (M^{tgt}_t)_{i,j} & \text{if } A(j) = \text{None}, \\
        (M_t)_{i,A(j)} & \text{otherwise}.
    \end{cases}
    \label{eq:7}
\end{equation}


\paragraph{\textbf{Local Attention Blends}}
Beyond global attention, we incorporate a blending mechanism suggested by~\citet{hertz2022prompt} and ~\citet{mokady2023null}. This method selectively integrates and maintains certain semantics by using target blend words $w^{tgt}$, which are words in the target
prompt whose semantics need to be added; and source blend words $w^{src}$, which are words in the source prompt whose semantics need to be preserved. At each denoising step $t$, the mechanism operates on the target latent $\vz^{tgt}_t$ as follows:
\begin{align}
    m_{\text{tgt}} &= \text{Threshold}\left[M^{tgt}_t(w_{\text{tgt}}), k_{\text{tgt}}\right], \\
    m_{\text{src}} &= \text{Threshold}\left[M^{src}_t(w_{\text{src}}), k_{\text{src}}\right],\\
    \vz^{tgt}_t &= (1-m^{tgt}+m^{src}) \odot \vz^{src}_t + (m^{tgt}-m^{src}) \odot \vz^{tgt}_t
    \label{eq:8-10}
\end{align}

\noindent where $m_{\text{src}}$ and $m_{\text{tgt}}$ are binary masks obtained by calibrating the aggregated attention maps $M^{src}_t(w^{src})$ and $M^{tgt}_t(w^{tgt})$ with threshold parameters $k_{src}$ and $k_{tgt}$, using the following threshold function:

\begin{equation}
    \text{Threshold}(M,k) = 
    \begin{cases}
        1 & \text{if } M_{i,j} \geq k, \\
        0 & \text{if } M_{i,j} < k.
    \end{cases}
    \label{eq:11}
\end{equation}

For simplicity, we define the process of local editing in Equation~\ref{eq:8-10} as:
\begin{equation}
    \vz^{tgt}_t = \text{LocalEdit}(\vz^{src}_t,\vz^{tgt}_t,M^{src}_t,M^{tgt}_t,w_{\text{src}},w_{\text{tgt}})
    \label{eq:12}
\end{equation}


\paragraph{\textbf{Scheduling Cross-Attention Control}}
Implementing cross-attention control throughout the entire sampling schedule can cause excessive focus on music structures, hindering the ability to incorporate intended changes. In contrast, applying it only during the early stages allows for creative flexibility while still preserving structural integrity.
Therefore, we limit cross-attention to the initial phases up to a cutoff point $\tau_c$. This moderation allows us to effectively capture the nuances and intended changes in musical compositions. The scheduling approach is defined as follows:
\begin{equation}
\begin{aligned}
    &\text{CrossEdit}(M^{src}, M^{tgt}, t) = \\
    &\begin{cases}
        \text{Refine}(M^{src}_t, M^{tgt}_t, t) & \text{if } t \geq \tau_c, \\
        M^{tgt}_t & \text{if } t < \tau_c.
    \end{cases}
\end{aligned}
\label{eq:13}
\end{equation}

\subsubsection{Mutual Self-Attention Control}
\label{4.2.2}

We diverge from the conventional use of cross-attention mechanisms and instead draw inspiration from the MasaCtrl technique~\citep{cao2023masactrl} to refine music structure through self-attention queries. 
These queries adeptly navigate through non-rigid musical transformations, aligning with the designated musical theme or instrument (target prompt). 
The process begins by sketching the foundational musical theme using the target's self-attention components—$Q^{tgt}$, $K^{tgt}$, and $V^{tgt}$. This is followed by enriching the theme with elements resembling the thematic content from the source ($K^{src}$, $V^{src}$), steered by $Q^{tgt}$. 
However, applying this attentive modulation uniformly over all processing layers and through every denoising step might result in a composition that excessively mirrors the source. Consequently, inspired by MasaCtrl, our proposed solution selectively employs mutual self-attention in the decoder portion of our music editing U-Net, initiated after a defined number of denoising iterations.

\paragraph{\textbf{Scheduling Mutual Self-Attention Control}} The application of the nuanced mutual self-attention control is meticulously planned, beginning at a specific denoising step $S$ and extending beyond a designated layer index $L$. The strength and influence of this control mechanism are designed as follows. 


\begin{equation}
\begin{aligned}
    &\text{SelfEdit}({Q^{src},K^{src},V^{src}}, {Q^{tgt},K^{tgt},V^{tgt}}, t) =\\ &\begin{cases}
        {Q^{src},K^{src},V^{src}} & \text{if } t \geq S \text{ and } l \geq L, \\
        {Q^{tgt},K^{src},V^{src}} & \text{otherwise}\\
    \end{cases}
    \label{eq:14}
\end{aligned}
\end{equation}

In this framework, $S$ serves
as a temporal threshold while $L$ tailors
the musical output towards the intended artistic direction.


\subsubsection{Harmonic Branch Integretion}
\label{4.2.3}

We hypothesize that a simple sequential combination of cross-attention control and mutual self-attention control may yield sub-optimal results within the original dual-branch framework. This approach is particularly ineffective in refining global attention, as depicted in Figure~\ref{fig:framework}(d). Our empirical validation, presented in Table~\ref{tab:3} (HAC vs w/o HB), supports this hypothesis.
To address this issue, we introduce an additional latent \textbf{harmonic branch}, which serves as an intermediate branch to host the desired composition and structural information in the target music. 
The unified framework of Harmonized Attention Control is detailed in Algorithm~\ref{alg:1}. During each forward step of the diffusion process, we start with mutual self-attention control on $z^{src}$ and $z^{tgt}$ and assign the output to the harmonic branch latent $z^{har}$. This latent lays the formal structure of the target music. Next, cross-attention control is applied on $M^{har}$ and $M^{tgt}$ to refine the semantic information for $M^{tgt}$. As illustrated in Figure~\ref{fig:attention}, the harmonic branch output $z^{har}_0$ reflects the requested non-rigid changes (e.g.,
``rapid''), while preserving the rigid content semantics (e.g., ``guitar''). The target branch output $z^{tgt}_0$ builds upon the structural layout of the  $z^{har}$ while
reflecting the requested rigid changes (e.g., ``violin'').

\begin{algorithm}[t]
\centering
\caption{Harmonized Attention Control in one DDIM Forward Process}
\begin{algorithmic}[1]
    \STATE \textbf{Input:}  A source prompt $\gP$, a target prompt $\gP^*$, a source audio latent $\vz_0$, denoising network $\epsilon_{\theta}(\cdot, \cdot, \cdot)$, current time step $\tau$, source and target blend words $w_{\text{src}}, w_{\text{tgt}}$, input latents $z^{\text{src}}_{\tau}, z^{\text{tgt}}_{\tau}, z^{\text{har}}_{\tau}$.
    \STATE $\epsilon_{\text{src}}, \{Q^{\text{src}}, K^{\text{src}}, V^{\text{src}}\}, M^{\text{src}} = \epsilon_{\theta}(z^{\text{src}}_{\tau}, \tau, c_{\text{src}})$
    \STATE $\epsilon_{\text{tgt}}, \{Q^{\text{tgt}}, K^{\text{tgt}}, V^{\text{tgt}}\}, M^{\text{tgt}} = \epsilon_{\theta}(z^{\text{tgt}}_{\tau}, \tau, c_{\text{tgt}})$
    \STATE $\{Q^{\text{har}}, K^{\text{har}}, V^{\text{har}}\} \newline= \text{SelfEdit}(\{Q^{\text{src}}, K^{\text{src}}, V^{\text{src}}\}, \{Q^{\text{tgt}}, K^{\text{tgt}}, V^{\text{tgt}}\}, \tau )$
    \STATE $\epsilon_{\text{har}}, M^{\text{har}} = \epsilon_{\theta}(z^{\text{har}}_{\tau}, \tau, c_{\text{src}}; \{Q^{\text{har}}, K^{\text{har}}, V^{\text{har}}\})$
    \STATE $\hat{M}^{\text{tgt}} = \text{CrossEdit}(M^{\text{har}}, M^{\text{tgt}}, \tau )$
    \STATE $\hat{\epsilon_{\text{tgt}}} = \epsilon_{\theta}(z^{\text{tgt}}_{\tau}, \tau, c_{\text{tgt}}; \hat{M}^{\text{tgt}})$
    \STATE $z^{\text{src}}_{\tau-1}, z^{\text{tgt}}_{\tau-1}, z^{\text{har}}_{\tau-1} \newline= \text{Sample}([z^{\text{src}}_{\tau}, z^{\text{tgt}}_{\tau}, z^{\text{har}}_{\tau}],[\epsilon_{\text{src}},\epsilon_{c_{\text{tgt}}},\epsilon_{\text{har}}], \tau )$
    \STATE $\vz^{tgt}_{\tau-1}$ = $\text{LocalEdit}(\vz^{src}_{\tau-1},\vz^{tgt}_{\tau-1},M^{src}_{\tau-1},M^{tgt}_t,w_{src},w_{tgt})$
    \STATE \textbf{Output:} $z^{\text{src}}_{\tau-1}, z^{\text{tgt}}_{\tau-1}, z^{\text{har}}_{\tau-1}$
\end{algorithmic}
\label{alg:1}
\end{algorithm}

\subsection{Disentangled Inversion Technique}
\label{4.3}
Using DDIM inversion without classifier-free guidance yields an easily modifiable but imprecise approximation of the original audio signal.  
Increasing the classifier-free guidance scale enhances editability, but sacrifices reconstruction accuracy due to latent code deviation during editing.

In order to address these limitations, we propose \textbf{Disentangled Inversion Technique} to disentangle the diffusion process into three branches: the source branch, the harmonic branch, and the target branch, with the detailed algorithm outlined in Algorithm~\ref{alg:2}.  This decoupling is designed to unleash the capabilities of each branch separately. For the source branch, we implement a targeted correction mechanism. By reintegrating the distance $z^*_t - z^{src}_t$ into $z^{src}_t$, we directly mitigate the deviation of the pathway. This straightforward adjustment effectively rectifies the path and minimizes the accumulated errors introduced by both DDIM inversion and classifier-free guidance, thereby enhancing consistency in the reconstructed audio. 
On the other hand, the target branch and harmonic branch are left unmodified to fully leverage the innate capabilities of diffusion models in generating the desired target audio, 
thereby ensuring the fidelity and integrity of the generated audio. Effectiveness of Disentangled Inversion Technique are verified and discussed in Section~\ref{5.4.2}. 

Typical diffusion-based editing~\citep{han2023improving,miyake2023negativeprompt} involves two parts: an inversion process to obtain the diffusion space of the audio, and a forward process to perform editing on the diffusion space. In contrast, \textbf{our Disentangled Inversion Technique can be plug-and-played into the forward process and rectifies the deviation path step by step}. Specifically, Disentangled Inversion first computes the difference between $\vz^*_{t-1}$ and $\vz^{src}_{t-1}$, then adds the difference back to $\vz^{src}_{t-1}$ in DDIM forward. We only add the difference of the source prompt in latent space and update $\vz^{src}_{t-1}$, which is the key to retaining the editability of the latent space of the target prompt.

\begin{algorithm}[htbp]
    \centering
    \caption{Disentangled Inversion Technique}
    \begin{algorithmic}[1]
     \STATE \textbf{Input}: A source prompt $\gP$, a target prompt $\gP^*$, a source audio latent $\vz_0$, and guidance scale $\omega$.
     \STATE \textbf{Output}: A edited audio latent $\vz^{tgt}_0$.
     \STATE Compute the intermediate results $\vz^*_T,...,\vz^*_1$ using
        DDIM inversion over $\vz_0$.
     \STATE Initialize $\vz^{src}_T \leftarrow \vz^*_T$, $\vz^{tgt}_T \leftarrow \vz^*_T$, $\vz^{har}_T \leftarrow \vz^*_T$.
    \FOR{$t=T$ to $1$}
    \STATE $[\vd^{src}_{t-1},\vd^{tgt}_{t-1},\vd^{har}_{t-1}] \leftarrow \vz^*_{t-1}$ \newline - DDIM\_Forward$(\vz^{src}_t,t,[\gP,\gP^*,\gP],\omega)$
    \STATE $\vz^{src}_{t-1} \leftarrow$ DDIM\_Forward($\vz^{src}_t,t,[\gP,\gP^*,\gP],\omega$) + $[\vd^{src}_{t-1},\vzero,\vzero]$
    \STATE $\vz^{har}_{t-1} \leftarrow$ DDIM\_Forward($\vz^{har}_t,t,[\gP,\gP^*,\gP],\omega$) + $[\vd^{src}_{t-1},\vzero,\vzero]$
    \STATE $\vz^{tgt}_{t-1} \leftarrow $ DDIM\_Forward($\vz^{tgt}_t,t,[\gP,\gP^*,\gP],\omega$) + $[\vd^{src}_{t-1},\vzero,\vzero]$
    \ENDFOR
    \STATE \textbf{Return:} $\vz^{tgt}_0$
    \end{algorithmic}
    \label{alg:2}
\end{algorithm}

%


\section{Experiments}
\begin{table*}[t]
    \centering
    \caption{Performance comparison of objective and subjective zero-shot music editing on ZoME-Bench (fixed length) and MusicDelta (variable length). The evaluation focuses on content preservation (SD, LPAPS, FAD, Chroma, MOS-P) and edit fidelity (CLAP Score, Accuracy, MOS-Q). Results include mean and standard deviation for both edit fidelity (MOS-Q) and content preservation (MOS-P). Metrics achieving the best scores are highlighted in bold.}
    \resizebox{0.9\linewidth}{!}{
    \begin{tabular}{c|cccccc|cc}
    \toprule
    \multirow{2}{*}{\bfseries Method} & \multicolumn{6}{c}{\bfseries Objective Metrics} & \multicolumn{2}{c}{\bfseries Subjective Metrics} \\
    & \bfseries $\text{SD}_{\times10^3}\downarrow$  & \bfseries LPAPS$\downarrow$ & \bfseries FAD$\downarrow$ &  \bfseries Chroma$\uparrow$ &  \bfseries CLAP Score$\uparrow$  & \bfseries $\text{Accuracy}\uparrow$ &\bfseries MOS-Q$\uparrow$ & \bfseries MOS-P$\uparrow$ \\
    \midrule
    \multicolumn{9}{c}{\textbf{Fixed Length Comparisons on ZoME-Bench}} \\
    \midrule
    AudioLDM 2 & 23.86  & 0.21 & 10.36  & 0.51 & 0.58 & 0.51& 73.48$\pm$0.92 & 70.12$\pm$1.23 \\
    MusicGen & 23.39  & 0.21 & 6.63 & 0.50& 0.59  & 0.52 & 75.46$\pm$1.19 & 71.28$\pm$0.99  \\
    SDEdit & 25.87 & 0.22 & 12.18  & 0.54& 0.40 & 0.35  & 69.38$\pm$1.47 & 66.23$\pm$1.38  \\
    DDIM Inversion & 22.52  & 0.21 & 9.51  & 0.57  & 0.49  & 0.44 & 73.10$\pm$1.24 & 73.38$\pm$1.16    \\
    MusicMagus & 16.23 & 0.19 & 5.15 & 0.62  & 0.55 &0.45& 75.12$\pm$1.08 & 74.34$\pm$1.17     \\
    DDPM-Friendly & 18.30 & 0.19 & 5.16  & 0.68 & 0.54  & 0.43 & 75.27$\pm$1.20 & 73.86$\pm$1.04  \\
    \bfseries MEDIC & \bfseries11.97  & \bfseries 0.15 & \bfseries 2.49 & \bfseries 0.73 & \bfseries 0.61  & \bfseries 0.59 & \bfseries 79.81$\pm$0.93 & \bfseries 77.29$\pm$0.88 \\
    \midrule
    \multicolumn{9}{c}{\textbf{Variable Length Comparisons On MusicDelta}} \\
    \midrule
    AudioLDM 2 & 24.40 & 0.22 & 7.07 & 0.51 & 0.44  & 0.48 & 66.37$\pm$1.11 & 64.28$\pm$1.20 \\
    MusicGen & 27.71& 0.23 & 7.70 & \bfseries 0.56& 0.46 & 0.34  & 67.41$\pm$1.35 & 63.76$\pm$1.29  \\
    SDEdit &28.12 &0.24 & 13.21  & 0.53& 0.24   & 0.38& 62.80$\pm$1.45 & 62.18$\pm$1.36  \\
    DDIM Inversion &23.5 & 0.21 & 10.12 & 0.52 & 0.27  & 0.41& 65.94$\pm$1.18 & 65.73$\pm$1.21    \\
    MusicMagus & 25.6&0.22 & 7.13 & 0.53 & 0.43   & 0.45&67.45$\pm$1.22 & 67.12$\pm$1.27     \\
    DDPM-Friendly& 21.53  &0.23 & 6.68 & 0.53 & 0.30  & 0.38& 66.34$\pm$1.03 & 67.28$\pm$1.10  \\
    \bfseries MEDIC & \bfseries 19.5  & \bfseries 0.20 & \bfseries 6.58 & 0.54 & \bfseries 0.51  & \bfseries 0.57&\bfseries 71.62$\pm$ 1.06 & \bfseries 70.18$\pm$0.97 \\
    \bottomrule
    \end{tabular}
    }
    
    \label{tab:1}
    \end{table*}

\subsection{Experimental Setup}
\label{subsec:exp_setup}
\noindent \textbf{Datasets.} %
To address the absence of a standardized benchmark for inversion and editing techniques and to systematically validate our proposed method as a plug-and-play strategy for music editing and compare with existing zero-shot music editing methods, we construct a new benchmark \textbf{ZoME-Bench (Zero-shot Music Editing Benchmark)}. 
ZoME-Bench comprises 1,100 audio samples, which are selected from MusicCaps~\citep{agostinelli2023musiclm} ZoME-Bench covers 10 different editing types that include both rigid and non-rigid modifications, detailed in supplementary material.
Each sample is accompanied by its corresponding source prompt, target prompt, human instruction, and source audio.

We evaluate different editing methods using the pre-trained AudioLDM 2 model~\citep{liu2023audioldm2} with 200 inference steps, on NVIDIA A800 GPU. The setting of all baselines follows that of ~\citet{manor2024zero}. All baseline editing methods are first evaluated on ZoME-Bench across all editing types of 10 seconds duration. We further use the commonly used \textbf{MusicDelta} subset of the MedleyDB~\citep{bittner2014medleydb} dataset for \textbf{variable length} performance comparisons. MusicDelta comprises 34 musical excerpts in varying styles and in lengths \textit{ranging from 20 seconds to 5 minutes}.

\noindent
\textbf{Evaluation Metrics.} Our comprehensive evaluation involves both objective and subjective metrics to assess essential content preservation, text-audio alignment fidelity, and audio quality. (1) \textbf{Objective Metrics:} We use a variety of objective metrics to measure different aspects of audio editing effectiveness. Structure Distance (SD)~\citep{ju2023direct}, CLAP Score (Contrastive Language-Audio Pretraining)~\citep{elizalde2023clap}, LPAPS (Learned Perceptual Audio Patch Similarity)~\citep{iashin2021taming,paissan2023audio}, Chroma (Chromagram Similarity), FAD (Fréchet Audio Distance)~\citep{kilgour2018fad}, and Accuracy are included in our evaluation framework. (2) \textbf{Subjective Metrics:} For subjective evaluation, we conduct crowd-sourced human assessments using the Mean Opinion Score (MOS). This metric is used to evaluate both edit fidelity (MOS-Q) and content preservation (MOS-P).



\subsection{Zero-shot Music Editing Results}
\label{6.3}


We present a comparative study of our MEDIC with DIC against several established music generation and editing baselines. We group these baselines into generation-based methods, including AudioLDM 2~\citep{liu2023audioldm2} and MusicGen~\citep{copet2023simple}, and inversion-based methods, including SDEdit~\citep{liu2023audioldm}, DDIM Inversion~\cite{ho2020denoising}, MusicMagus~\citep{zhang2024musicmagus}, and DDPM-Friendly~\citep{manor2024zero}.

\noindent
\textbf{Fixed Length Comparisons.} 
We evaluate the generated audio samples on the ZoME-Bench test set with a fixed length, focusing on the key aspects of content preservation and edit fidelity. As exhibited in Table~\ref{tab:1}, the results yield the following insights: (1) \textbf{Our MEDIC substantially outperforms both generation inversion-based models in terms of edit fidelity and content preservation in both objective and subjective metrics, demonstrating its effectiveness in addressing complex editing tasks}. (2)
While DDPM-Friendly and MusicMagus improve content preservation (higher MOS-P), they lag in text-audio alignment and editing precision, indicated by their lower CLAP Scores and Accuracy. In contrast, MEDIC achieves consistently better CLAP, Accuracy, LPAPS, FAD, and Chroma, demonstrating superior alignment with target prompts as well as overall musical similarity.

\noindent
\textbf{Variable Length Comparisons}  
We further evaluate MEDIC against the baseline methods in a \textit{variable length} setting on the MusicDelta dataset. Table~\ref{tab:1} highlights the following insights: (1) MEDIC consistently outperforms all baselines across all objective and subjective metrics, demonstrating strong robustness and adaptability for editing longer audio segments. (2) Inversion-based baselines suffer a notable drop in edit fidelity, as seen in their lower Accuracy and CLAP scores, likely due to error accumulation and insufficient attention control. In contrast, MEDIC achieves the highest CLAP and Accuracy, confirming its ability to deliver precise and well-aligned edits even on audio longer than 20 seconds.

\noindent
\textbf{Fine-grained Comparisons on ZoME-Bench.}
We further compare performance across different editing types using FAD and CLAP scores (Figure~\ref{fig:spider}):
(1) MEDIC consistently outperforms all baselines for both rigid and non-rigid editing tasks.
(2) Baselines can handle some rigid edits, but perform poorly on non-rigid manipulations such as “Change Genre” and “Change Melody”.

\begin{figure*}[t]
    \centering
    \vspace{-2mm}
    \includegraphics[width=.99\textwidth]{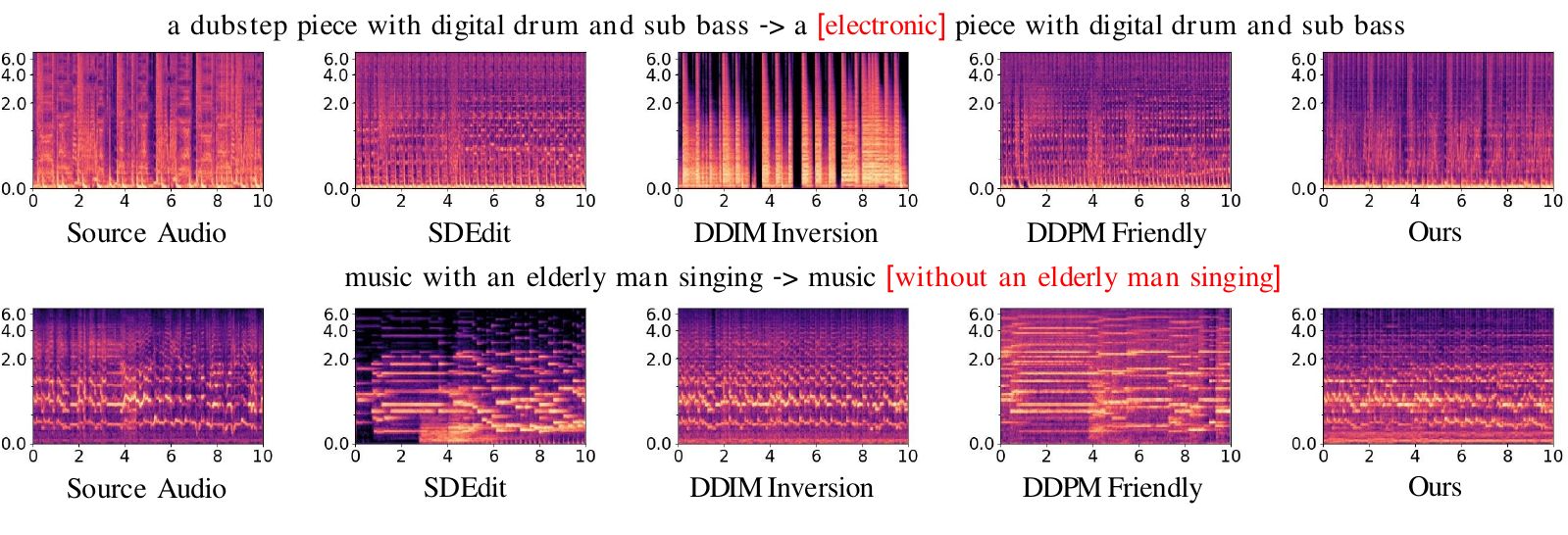}
    \caption{Visualizations of the mel-spectrograms of the source audio and the edited audio by different editing methods.}
    \vspace{-2mm}
    \label{fig:visual}
\end{figure*}

\subsection{Ablation Study and Visualization}
\label{6.4}


\noindent
\textbf{Effect of Attention Control Methods.}
To validate our attention control designs, we conduct ablation studies on the following configurations: Remove Mutual Self-Attention Control (w/o MSA Control), Remove Cross-Attention Control (w/o CA Control), and Remove Harmonic Branch (w/o Harmonic Branch). Table~\ref{tab:3} reveals that: 
(1) Both cross-attention and mutual self-attention controls significantly enhance editing performance. Their presence leads to improvements across all metrics, demonstrating their crucial role in achieving higher content preservation and edit fidelity.
(2) While the naive combination of mutual self-attention and cross-attention control improves metrics such as LPAPS, FAD, and Chroma, it results in sub-optimal outcomes without the inclusion of the Harmonic Branch. This missing component reduces overall coherence and refinement, emphasizing the critical role of the Harmonic Branch in augmenting attention control mechanisms to optimize both content integrity and edit fidelity.

\begin{table}[htbp]
\centering
\caption{Effect of Mutual Self-Attention control (MSA), Cross-Attention (CA) control, and Harmonic Branch (HB).}
\vspace{1mm}
\resizebox{0.95\linewidth}{!}{
\begin{tabular}{c|cccccc}
\toprule
    \bfseries Method & \bfseries $\text{SD}_{\times10^3}\downarrow$  & \bfseries LPAPS$\downarrow$ & \bfseries FAD$\downarrow$  & \bfseries Chroma$\uparrow$ & \bfseries CLAP$\uparrow$ & \bfseries $\text{Accuracy}\uparrow$  \\
\midrule
\bfseries HAC & \bfseries 11.97  & \bfseries 0.15 & \bfseries 2.49   & \bfseries 0.73 & \bfseries 0.61 & \bfseries 0.59 \\
\midrule
w/o MSA Ctrl & 14.13 & 0.19 & 2.75  & 0.57  &  0.56&  0.45  \\
w/o CA Ctrl & 13.78   & 0.18 & 2.50 & 0.63  & 0.58& 0.48  \\
w/o HB & 12.75& 0.16 & 2.59  & 0.66 & 0.59 & 0.51 \\
\bottomrule
\end{tabular}
}
\label{tab:3}
\end{table}

\label{5.4.2}
\noindent
\textbf{Effectiveness of Disentangled Inversion Technique.}
We assess the soundness of Algorithm~\ref{alg:2} and the effectiveness of disentanglement of triple branches. As shown in Table~\ref{tab:4}, 1) When employing $[\vd_{src},\vzero,\vzero]$—the configuration used in MEDIC—results are notably strong across several metrics. MEDIC achieves the highest Accuracy (0.59), CLAP score (0.61), and Chroma (0.73), demonstrating its balanced approach in effectively preserving content fidelity while maintaining strong alignment with the target prompt. 2) Configurations that incorporate additional distances, such as $[\vd_{src},\vd_{src},\vzero]$ and $[\vd_{src},\vd_{har},\vzero]$, show a notable decline in performance across most metrics.
3)Interestingly, the configuration $[\vd_{src},\vzero,\vd_{tgt}]$ achieves the lowest values for LPAPS and FAD, suggesting potential improvements in audio similarity. However, this setup seems to markedly reduce overall text-audio alignment and accuracy, indicating a trade-off where structural precision comes at the expense of editorial coherence.

\begin{table}[htbp]
\centering
\vspace{1mm}
\caption{Ablation study of Disentangled Inversion Technique. $[\cdot,\cdot,\cdot]$ denotes adding the distance (line 6 in Algorithm~\ref{alg:2}). MSE is the mean square error loss between the edited audio features and source audio features. $[\vd_{src},\vzero,\vzero]$ is used in MEDIC.}
\resizebox{0.95\linewidth}{!}{
\begin{tabular}{c|cccccc}
\toprule
    \bfseries Distance & \bfseries $\text{SD}_{\times10^3}\downarrow$ & \bfseries LPAPS$\downarrow$ & \bfseries FAD$\downarrow$& \bfseries Chroma$\uparrow$ & \bfseries CLAP$\uparrow$  & \bfseries $\text{Accuracy}\uparrow$ \\
\midrule
$[\vd_{src},\vzero,\vzero]$ &11.97 &  0.15 &  2.49   &  \bfseries 0.73 & \bfseries 0.61&  \bfseries 0.59\\
\midrule
$[\vd_{src},\vd_{src},\vzero]$ &14.28 & 0.17 & 2.64   & 0.53 & 0.57& 0.47 \\
$[\vd_{src},\vzero,\vd_{src}]$ &13.17 & 0.17 & 2.55  & 0.47 & 0.58  & 0.52 \\
$[\vd_{src},\vd_{tgt},\vzero]$ &11.24  & 0.15 & 2.44  & 0.52 & 0.56 & 0.52 \\
$[\vd_{src},\vzero,\vd_{tgt}]$ & \bfseries 11.12  & \bfseries 0.14 & \bfseries 2.41  & 0.55 & 0.56 & 0.32\\
$[\vd_{src},\vd_{har},\vzero]$ &37.51 & 0.27 & 2.72   & 0.32 & 0.28 & 0.31\\
$[\vzero,\vzero,\vzero]$ & 12.65 &  0.16 &  2.54   & 0.47 & 0.57 & 0.47  \\
\bottomrule
\end{tabular}
}
\label{tab:4}
\end{table}

\noindent \textbf{Visualization.}
To complement our quantitative findings, we present a qualitative comparison in Figure~\ref{fig:visual}. Methods such as SDEdit and inversion-based techniques often struggle to balance high editability and preserve melodic content and harmonic structure. In contrast, MEDIC performs better in precise music editing while preserving structural integrity. We provide additional qualitative results for all editing categories in supplementary material, demonstrating the superiority of our approach.
\section{Conclusion}
\label{7}
In this paper, 
we propose the Disentangled Inversion Control to support both rigid and non-rigid editing tasks and develop a zero-shot music editing framework MelodyEdit based on DIC. We add an intermediate harmonic branch to progressively integrate harmonic and melodic information in music by cross-attention control and mutual self-attention control. To counteract the accumulated errors caused by DDIM inversion and CFG, we introduce Disentangled Inversion to separate the diffusion process into triple branches and eliminate the latent discrepancy distance in the source branch. Extensive experiments demonstrate superiority of MelodyEdit on both fixed and variable length settings. We envisage that our work could serve as a basis for future zero-shot music editing studies. 
\begin{acks}
The research was supported by the Early Career Scheme (ECS-Hong Kong University of Science and Technology 22201322) and NSFC (No. 62206234) of Mainland China.
\end{acks}

\bibliographystyle{ACM-Reference-Format}
\bibliography{sample-base}
\newpage
\section*{Table of Contents}

\begin{enumerate}
    \item \textbf{Benchmark construction}
        \begin{enumerate}
            \item \textbf{General information}
            \item \textbf{Annotation Process}
            \item \textbf{Data Format}
        \end{enumerate}
    \item \textbf{Implementation Details}
        \begin{enumerate}
            \item \textbf{Metrics}
        \end{enumerate}
    \item \textbf{Quantitative Results}
    \item \textbf{Potential Negative Societal Impacts} 
    \item \textbf{Limitations}
    \item \textbf{Qualitative Results}
        \begin{enumerate}
            \item \textbf{Change Instrument}
            \item \textbf{Add Instrument}
            \item \textbf{Delete Instrument}
            \item \textbf{Change Genre}
            \item \textbf{Change Mood}
            \item \textbf{Change Rhythm}
            \item \textbf{Change Background}
            \item \textbf{Change Melody}
            \item \textbf{Extract Instrument}
        \end{enumerate}
    \item \textbf{safeguards}
\end{enumerate}
\section{Benchmark construction}
\label{B}
\subsection{General information}
\label{B.1}
Here are the details of our ZoME-Bench dataset (Zero-shot Music Editing Benchmark). This dataset contains 1,000 audio samples, selected from MusicCaps, with each sample being 10 seconds long and having a sample rate of 16k.

We refactor the original captions to express specific edits and divide them into 10 parts, each representing a different type of editing. A sample and details are shown in the following table~\ref{tab:my_label}. 

\begin{table*}[ht]
\centering
\renewcommand{\arraystretch}{1.5}
\begin{tabular}{|c|c|c|p{2.4cm}|p{2.4cm}|p{2.4cm}|}

\hline
\makecell{Editing\\ type id} & Editing type  & size & origin prompt & editing prompt& editing instruction\\
\hline
0 & change instrument & 131 & ambient acoustic [guitar] music &ambient acoustic [violin] music &change the instrument from guitar to violin \\
1 & add instrument & 139 & metal audio with a distortion guitar [and drums] &metal audio with a distortion guitar & add drums to the piece \\
2 & delete instrument & 133&an eerie tense instrumental featuring electronic drums [and synth keyboard]&an eerie tense instrumental featuring electronic drums & remove the synth keyboard\\
3 & change genre & 134 &a recording of a solo electric guitar playing [blues] licks&a recording of a solo electric guitar playing [rocks] licks & change the genre from blues to rock\\
4 & change mood& 100 &a recording featuring electric bass with an [upbeat] vibe &a recording featuring electric bass with an [melancholic] vibe & turn upbeat mood into melancholic mood\\
5 & change rhythm & 69 &a live ukulele performance featuring [fast] strumming and emotional melodies&a live ukulele performance featuring [slow] strumming and emotional melodies & change fast rhythm into slow one \\
6 & change background& 95 &female voices in unison with [acoustic] guitar&female voices in unison with [electric] guitar & switch acoustic guitar to electric guitar\\
7 & change melody & 121&this instrumental song features a [relaxing] melody&this instrumental song features a [cheerful] melody & change relaxing melody into cheerful melody \\
8 & extract instrument & 111 &a reggae rhythm recording with bongos [djembe congas acoustic drums and electric guitar]&a reggae rhythm recording with bongos & extract bongos from the recording\\
9 & random & 67 &/ &/ &/\\
\hline
\end{tabular}
\caption{Information of ZoME-Bench dataset}
\label{tab:my_label}
\end{table*}

\subsection{Annotation Process}
\label{B.2}
We rebuild our caption from captions for Musiccaps offered by ~\citep{agostinelli2023musiclm}. With the help of ChatGPT-4~\citep{openai2023gpt4}, we rebuild the caption with prompt as follows(take type ``change melody" as examples):

        {\textbf{Description}: ``There is a description of a Piece of music, Please judge whether the description has information of melody. If not, just answer ``Flase", else change its melody properly into the opposite one, just change the adjective and don't replace any instrument!
                "},{``blended\_word" is [origin melody, changed melody],
                ``emphasize" is [changed melody],
                ``blended\_word" and ``emphasize" are tuples.\\
                \textbf{Question}: (A mellow, passionate melody from a noisy electric guitar) \\
                 \textbf{Answer}:(``source\_prompt": ``A mellow, [passionate] melody from a noisy electric guitar",
                        ``editing\_prompt": ``A mellow, [soft] melody from a noisy electric guitar",
                        ``blended\_word": [``passionate melody", ``soft melody"],
                        ``emphasize": [``soft melody"])

                \noindent         
                \textbf{Question}: (A recording of solo harp music with a dreamy, relaxing melody.)\\
                \textbf{Answer}: (``source\_prompt": ``A recording of solo harp music with a dreamy, [relaxing] melody.",
                        ``editing\_prompt": ``A recording of solo harp music with a dreamy, [nervous] melody.",
                        ``blended\_word": [``relaxing melody",``nervous melody"],
                        ``emphasize" :[``nervous melody"])\\
                \textbf{Question}: (``A vintage, emotional song with mellow harmonized flute melody and soft wooden percussions")\\
                \textbf{Answer}: (``source\_prompt": ``A vintage, emotional song with [passionate] flute melody and soft wooden percussions.",
                        ``editing\_prompt": ``A vintage, emotional song with [harmonized] flute melody and soft wooden percussions.",
                        ``blended\_word": [``harmonized flute melody",``passionate flute melody"]),
                        ``emphasize" :[``passionate flute melody"])\\
            \textbf{Now we have
                Question}:(\{origin caption\}), Answer(?)"

In the same way, instructions are appended by prompt as follows (take type ``change melody" as examples):

\textbf{Description}: ``There are two descriptions of different pieces of music divided by \&, Please describe the difference\
                you need to give me the results in the following format:\
                Question: this instrumental song features a [relaxing] melody with a country feel accompanied by a guitar piano simple percussion and bass in a slow tempo \& this instrumental song features a [cheerful] melody with a country feel accompanied by a guitar piano simple percussion and bass in a slow tempo\\
                \textbf{Answer}: change relaxing melody  into cheerful melody \\
                \textbf{Question}: this song features acapella harmonies with a [high pitched] melody complemented by both high pitched female whistle tones and male low pitch tones \& this song features acapella harmonies with a [smooth] melody complemented by both high pitched female whistle tones and male low pitch tones \\
                \textbf{Answer}: turn a high pitched melody into smooth melody \\
                \textbf{Question}: a traditional and hopeful song with a harmonizing throaty male vocal and [dissonant] background melody from strings albeit presented in low quality \& a traditional and hopeful song with a harmonizing throaty male vocal and [harmonic] background melody from strings albeit presented in low quality \\
                \textbf{Answer}: change dissonant melody into harmonic melody \\
                \textbf{Now we have Question}: {[`source prompt']} \& {[`editing prompt']}, Answer(?)"

Through this method, supplemented by rounds of manual review, we ensure the quality of this benchmark.
 
\subsection{Data Format}
\label{B.3}
Taking the first piece as an example, we express our data in JSON format with six keys
\begin{lstlisting}[firstnumber=1,basicstyle=\ttfamily,breaklines=true]
{
    "000000000000": {
    "editing_prompt": "a live recording of ambient acoustic
    [violin] music",
    "source_prompt": "a live recording of ambient acoustic
    [guitar] music",
    "blended_word": "("guitar", "violin")",
    "emphasize": "("violin")",
    "audio_path": "wavs/MusicCaps_-4SYC2YgzL8.wav",
    "editing_type_id": "0",
    "editing_instruction": "change the instrument from guitar 
    to violin"
    }
}
\end{lstlisting}
``Editing\_prompt" refers to the edited caption, while ``source\_prompt" denotes the original caption. "Blended\_word" indicates the subject to be edited, and ``Emphasize" represents the word that should be highlighted. ``Editing\_instruction" provides a description of the editing process. Additionally, in the editing type ``delete instrument," we introduce another key, ``neg\_prompt", which helps reduce the likelihood of deleted instruments reappearing.

\section{Implementation Details}
\label{C}
For our evaluation, we utilize the pre-trained AudioLDM 2-Music model~\citep{liu2023audioldm2}. Our assessment employs a comprehensive set of metrics, namely CLAP, LPAPS, Structure Distance, and FAD. These metrics are calculated using the CLAP models available in the AudioLDM\_eval package, which is accessible at \url{https://github.com/haoheliu/audioldm_eval}. In line with the methodology described by ~\citet{manor2024zero}, we apply a forward guidance of 3 and a reverse guidance scale of 12 for DDPM inversion. For the DDIM inversion, the guidance scale is set to 5, while for SDEdit, we employ a guidance scale of 12. The forward guidance of MelodyEdit is 1 while the reverse scale is 5. We chose these values by exploring different guidance scales, as discussed in Appendix~\ref{D.1}. We conduct all experiments in NVIDIA 4090.


Our methodology is aligned with the protocol established by ~\citet{manor2024zero}, where we have adopted a forward guidance scale of 3 and a reverse guidance scale of 12 for DDPM inversion. In contrast, the DDIM inversion employs a guidance scale of 5, and SDEdit utilizes a guidance scale of 12. For Disentangled Inversion Control, we have determined the forward guidance to be 1 and the reverse scale to be 5. These specific guidance scale values are selected after extensive experimental exploration, the details of which are discussed in Appendix~\ref{D}.
\subsection{Metrics}\label{C.1}
\noindent
\textbf{Objective Metrics}
There are details about four metrics to evaluate the performance of our novel Disentangled Inversion Control framework:
(1) \textbf{CLAP Score}~\citep{elizalde2023clap}: This criterion evaluates the degree to which the output conforms to the specified target prompt.
(2) \textbf{Struture Distance}~\citep{ju2023direct}: Leveraging self-similarity of audio features to measure the structure distance between the source and edited audio.
(3) \textbf{LPAPS}~\citep{iashin2021taming,paissan2023audio}: An audio adaptation of the Learned Perceptual Image Patch Similarity (LPIPS)~\citep{zhang2018unreasonable}, this measure evaluates the consistency of the edited audio with the source audio.
(4) \textbf{FAD (Fréchet Audio Distance)}~\citep{kilgour2018fad}: Analogous to the FID used in image analysis, this metric calculates the distance between two distributions of audio signals.
(5) \textbf{Chroma (Chromagram Similarity)}~\citep{zhang2024musicmagus}: The average cosine similarity between the chromagrams of the original and edited music, which denotes the preservation of pitch contours and rhythm patterns in the music.
(6) \textbf{Accuracy}~\citep{chu2024qwen2}: The rate of successful editing judged by the Qwen model, calculated by constructing a question-answer (QA) pair, where the model’s output is compared against the expected result. The comparison assesses whether the model has made the correct edit.

\noindent
\textbf{Subjective Metrics}
To directly reflect the quality of the audio generated, we carry out MOS (Mean Opinion Score) tests. These tests involve scoring two aspects: MOS-Q, which assesses the edited quality of the audio, and MOS-P, which measures the content preservation of edited audio.

For assessing editing fidelity, the evaluators were specifically directed to “Does the natural language description align with the audio?” They were provided with both the audio and its corresponding caption. They were then asked to give their subjective rating (MOS-Q) on a 20-100 Likert scale.

To assess essential content preservation, human evaluators were presented with source audio, target audio, source prompt, and target prompt. They were then asked to answer the question, “To what extent does the target audio retain the essential features of the source audio, such as melody, instrumentation, and overall style?” The raters had to select one of the options: “completely,” “mostly,” or “somewhat,” using a 20-100 Likert scale for their response.

Our crowd-sourced subjective evaluation tests were conducted via Amazon Mechanical Turk where participants were paid \$8 hourly.
\section{Quantitative Results}
\label{D}
\noindent
\textbf{Analyses on Different CFG Scale}\label{D.1}
The lack of systematic experiments that determine the optimal combination of guidance scales for achieving the
best editing performance, and analysis of how these guidance scales affect the final consequence in both reconstruction and editing, we conduct this experiment to find the best scales.

\begin{table*}[t]
\centering
\setlength{\tabcolsep}{2mm}{
\begin{tabular}{c|c|c|ccc|c}
\toprule
\multicolumn{2}{c|}{\textbf{Guidance Scale}}           & \textbf{Structure}          & \multicolumn{3}{c|}{\textbf{Background Preservation}} & \textbf{CLIP Similariy} \\ \midrule
\textbf{Inverse}          & \textbf{Forward}            & $\textbf{Distance}_{\times10^3}\downarrow$ & \bfseries LPAPS$\downarrow$ & \bfseries FAD$\downarrow$ & \bfseries $\text{MSE}_{\times10^5}\downarrow$    & \textbf{CLAP Score} $\uparrow$    \\ \midrule
1   & 1   &8.56   &0.12 &1.17  &3.25	&0.51 \\
1   & 2.5 &11.97  &0.15 &2.49  &4.54	&0.61\\
1   & 5   &15.99  &0.17	&4.22  &6.07	&0.61\\
1   & 7.5 &15.99  &0.17	&4.22  &6.07	&0.59\\
2.5 & 1   &22.80  &0.20 &6.39  &8.65    &0.30\\
2.5 & 2.5 &14.24  &0.16 &2.50  &5.40    &0.46\\
2.5 & 5   &14.46  &0.16 &3.31  &5.49    &0.53\\
2.5 & 7.5 &15.51  &0.17 &3.94  &5.89	&0.53\\
5   & 1   &29.94  &0.24 &9.81  &11.36	&0.20\\
5   & 2.5 &29.16  &0.24	&9.11  &11.07	&0.22\\
5   & 5   &22.15  &0.20	&5.59  &8.40    &0.36\\
5   & 7.5 &17.57  &0.18	&5.57  &6.67	&0.48\\
7.5 & 1   &31.41  &0.25	&10.62 &11.92	&0.20\\
7.5 & 2.5 &31.05  &0.25	&10.14 &11.78	&0.20\\
7.5 & 5   &29.20  &0.24	&9.32  &11.08	&0.24\\
7.5 & 7.5 &24.16  &0.22	&7.33  &9.17    &0.34 \\
\bottomrule
\end{tabular}}
\vspace{1mm}
\caption{Ablation Studies on Different Guidance Scale}
\vspace{-2mm}
\label{tab:ablation_guidance}
\end{table*}

\noindent
\textbf{Inference Time}\label{D.2}
We compare the inference time of our method with baselines, and the results are compiled in Table~\ref{tab:time}. MelodyEdit achieves the comparative inference time with generation models and inversion techniques. We will make an attempt to reduce the inference time of zero-shot music editing in our future work.

\begin{table}[t]
\centering
\begin{tabular}{cc}
\toprule
    \bfseries Method & \bfseries \text{Inference Time} \\
\midrule
AudioLDM 2  &  42.5s\\
MusicGen & 83.3s \\
SDEdit  & 44.3s \\
DDIM Inversion  &  81.6s\\
MusicMagus  & 89.0s \\
DDPM-Friendly & 33.3s \\
MelodyEdit & 92.0s \\
\bottomrule
\end{tabular}
\vspace{2mm}
\caption{Inference Time across different methods.}
\label{tab:time}
\vspace{-4mm}
\end{table}

\section{Potential Negative Societal Impacts} 
\label{E}
MelodyEdit may also lead to potential negative societal impacts that are worthy of consideration. If the data sample of the training model is not diverse enough or biased, the AI-generated music may be overly biased toward one style or element, limiting the diversity of the music and causing discrimination.MelodyEdit could be used to create fake audio content, such as faking someone's voice or creating fake musical compositions, posing the risk of fraud and impersonation. Hopefully, all these issues could be taken into consideration when taking the model for real use to avoid ethical issues.

\section{Limitations}
\label{F}
In spite of the remarkable outcome of our method, due to the limitation of the generation model we used, we are incapable of instigating a profound change.

Due to the numerous steps it requires (T=200), the duration of computing distance is quite long. Thus, we will implement a more powerful text-to-music generation model to support better editing, while trying to use a consistency model or flow-matching model to achieve high-quality and fast music generation in future work. We will make an attempt to edit more interesting and complex music tasks in the future.

\section{Qualitative Results}
\label{G}
For each type in ZoME-Bench, We provide samples to observe the capability of MelodyEdit intuitively.

\subsection{Change Instrument}
\label{G.1}
In Figure~\ref{fig:type0}, we show the capability of MelodyEdit to change the instrument. Here we edit the ground truth music piece with the source prompt ``a live recording of ambient acoustic [guitar] music" and editing prompt ``a live recording of ambient acoustic [violin] music". The difference in instruments can be observed in the Mel-spectrum.

\begin{figure*}[htbp]
\centering
\includegraphics[width=0.8\textwidth, height=0.1\textheight]{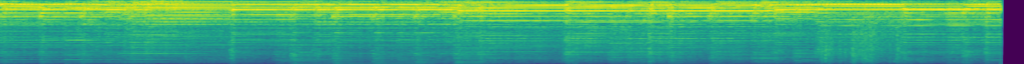}
\\source: a live recording of ambient acoustic [guitar] music\\
\includegraphics[width=0.8\textwidth, height=0.1\textheight]{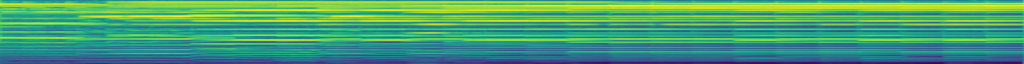} 
\\edit: a live recording of ambient acoustic [violin] music\\
\caption{Editing Type 0 :Change Instrument} 
\label{fig:type0}
\end{figure*}

\subsection{Add Instrument}
\label{G.2}
In Figure~\ref{fig:type1}, we show the capability of MelodyEdit to add more instruments. Here we edit the ground truth music piece with the source prompt ``a heavy metal instructional audio with a distortion guitar" and editing prompt ``a heavy metal instructional audio with a distortion guitar [and drums]". The appearance of the new instrument can be observed in the Mel-spectrum which presents a drum sound of high frequency.

\begin{figure*}[htbp]
\centering
\includegraphics[width=0.8\textwidth, height=0.1\textheight]{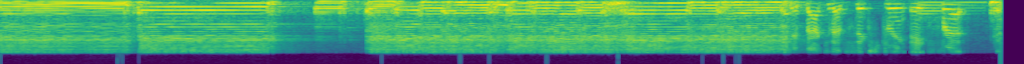}
\\source: a heavy metal instructional audio with a distortion guitar\\
\includegraphics[width=0.8\textwidth, height=0.1\textheight]{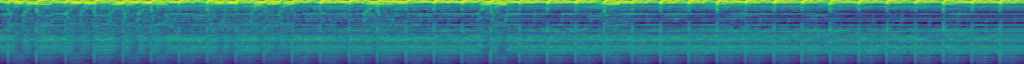} 
\\edit: a heavy metal instructional audio with a distortion guitar [and drums]\\
\caption{Editing Type 1 Add Instrument} 
\label{fig:type1}
\end{figure*}

\subsection{Delete Instrument}
\label{G.3}
In Figure~\ref{fig:type2}, we show the capability of MelodyEdit to delete instruments. Here we edit the ground truth music piece with the source prompt ``a lively ska instrumental featuring keyboard trumpets bass [and percussion] with a groovy mood" and the editing prompt ``a lively ska instrumental featuring keyboard trumpets and bass with a groovy mood". The vanishing of the instrument can be observed in the Mel-spectrum.

\begin{figure*}[htbp]
\centering
\includegraphics[width=0.8\textwidth, height=0.1\textheight]{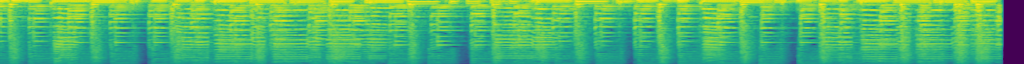}
\\source: a lively ska instrumental featuring keyboard trumpets bass [and percussion] with a groovy mood\\
\includegraphics[width=0.8\textwidth, height=0.1\textheight]{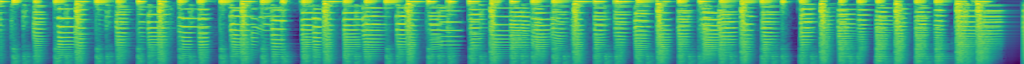} 
\\edit: a lively ska instrumental featuring keyboard trumpets and bass with a groovy mood\\
\caption{Editing Type 2 Delete Instrument}
\label{fig:type2}
\end{figure*}

\subsection{Change Genre}
\label{G.4}
In Figure~\ref{fig:type3}, we show the capability of MelodyEdit to change the genre of a music piece. Here we edit the ground truth music piece with the source prompt ``a recording of a solo electric guitar playing [blues] licks" and the editing prompt ``a recording of a solo electric guitar playing [rock] licks". The obvious difference in genre can be observed in the Mel-spectrum.

\begin{figure*}[htbp]
\centering
\includegraphics[width=0.8\textwidth, height=0.1\textheight]{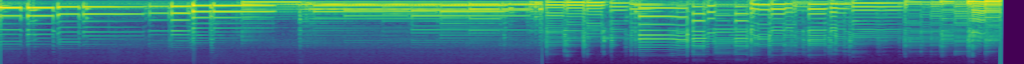}
\\source: a recording of a solo electric guitar playing [blues] licks\\
\includegraphics[width=0.8\textwidth, height=0.1\textheight]{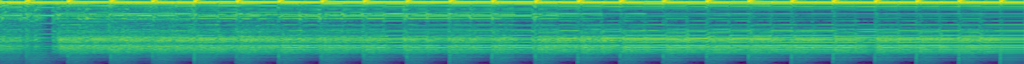} 
\\edit: a recording of a solo electric guitar playing [rock] licks\\
\caption{Editing Type 3 Change Genre} 
\label{fig:type3}
\end{figure*}

\subsection{Change Mood}\
\label{G.5}
Mood is an important attribute of music. In Figure~\ref{fig:type4}, we show the capability of MelodyEdit to change the mood of a music piece. Here we edit the ground truth music piece with the source prompt ``a recording of [aggressive] electronic and video game music with synthesizer arrangements" and editing prompt ``a recording of [peaceful] electronic and video game music with synthesizer arrangements". The change of mood can be observed in the Mel-spectrum.

\begin{figure*}[htbp]
\centering
\includegraphics[width=0.8\textwidth, height=0.1\textheight]{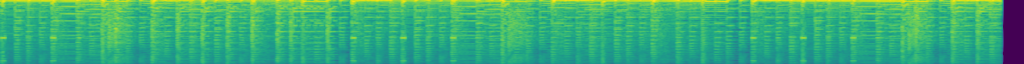}
\\source: a recording of [aggressive] electronic and video game music with synthesizer arrangements\\
\includegraphics[width=0.8\textwidth, height=0.1\textheight]{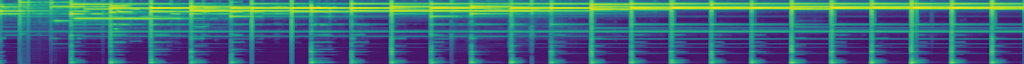} 
\\edit: a recording of [peaceful] electronic and video game music with synthesizer arrangements\\
\caption{Editing Type 4 Change Mood} 
\label{fig:type4}
\end{figure*}

\subsection{Change Rhythm}
\label{G.6}
Rhythm represents the speed of the music. In Figure~\ref{fig:type5}, we show the capability of MelodyEdit to change the Rhythm of a music piece. Here we edit the ground truth music piece with the source prompt ``a [slow] tempo ukelele tuning recording with static" and the editing prompt ``a [fast] tempo ukelele tuning recording with static". The change of Rhythm can be observed in the Mel-spectrum. The edited Mel-spectrum is much more intensive.

\begin{figure*}[htbp]
\centering
\includegraphics[width=0.8\textwidth, height=0.1\textheight]{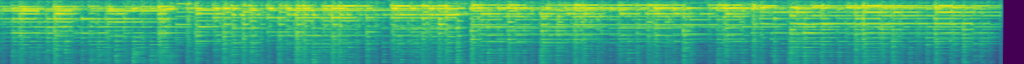}
\\source: a [slow] tempo ukelele tuning recording with static\\
\includegraphics[width=0.8\textwidth, height=0.1\textheight]{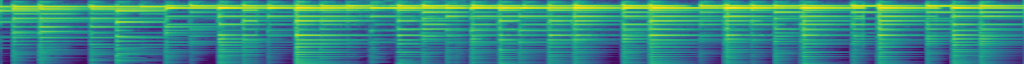} 
\\edit: a [fast] tempo ukelele tuning recording with static\\
\caption{Editing Type 5 Change Rhythm} 
\label{fig:type5}
\end{figure*}

\subsection{Change Background}
\label{G.7}
In Figure~\ref{fig:type6}, we show the capability of MelodyEdit to change the background of the instrument of a music piece. Here we edit the ground truth music piece with the source prompt ``an amateur ukulele recording with a [medium to uptempo] pace" and editing prompt ``an amateur ukulele recording with a [steady and rhythmic] pace". The change of instrument background can be observed in the Mel-spectrum.

\begin{figure*}[htbp]
\centering
\includegraphics[width=0.8\textwidth, height=0.1\textheight]{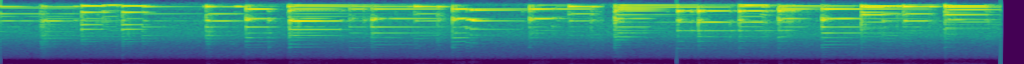}
\\source: an amateur ukulele recording with a [medium to uptempo] pace" and editing prompt \\
\includegraphics[width=0.8\textwidth, height=0.1\textheight]{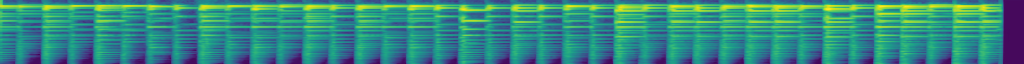} 
\\edit: an amateur ukulele recording with a [steady and rhythmic] pace\\
\caption{Editing Type 6 Change Background} 
\label{fig:type6}
\end{figure*}

\subsection{Change Melody}
\label{G.8}
In Figure~\ref{fig:type7}, we show the capability of MelodyEdit to change the melody of a music piece. Here we edit the ground truth music piece with the source prompt ``this instrumental song features a [relaxing] melody with a country feel accompanied by a guitar piano simple percussion and bass in a slow tempo" and editing prompt ``this instrumental song features a [cheerful] melody with a country feel accompanied by a guitar piano simple percussion and bass in a slow tempo". The change of Melody can be observed in the Mel-spectrum.

\begin{figure*}[htbp]
\centering
\includegraphics[width=0.8\textwidth, height=0.1\textheight]{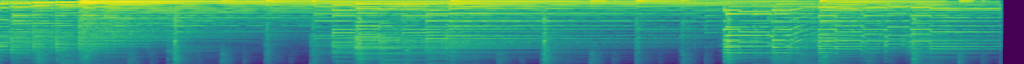}
\\source: this instrumental song features a [relaxing] melody with a country feel accompanied by a guitar piano simple percussion and bass in a slow tempo \\
\includegraphics[width=0.8\textwidth, height=0.1\textheight]{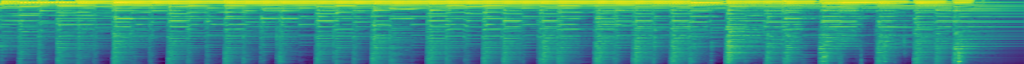} 
\\edit: this instrumental song features a [cheerful] melody with a country feel accompanied by a guitar piano simple percussion and bass in a slow tempo\\
\caption{Editing Type 7 Change Melody}
\label{fig:type7}
\end{figure*}

\subsection{Extract Instrument}
\label{G.9}
In Figure~\ref{fig:type8}, we show the capability of MelodyEdit to extract one certain instrument of a music piece. Here we edit the ground truth music piece with the source prompt ``a reggae rhythm recording with bongos [djembe congas acoustic drums and electric guitar]" and editing prompt ``a reggae rhythm recording with bongos". The change of instruments can be observed in the Mel-spectrum.

\begin{figure*}[htbp]
\centering
\includegraphics[width=0.8\textwidth, height=0.1\textheight]{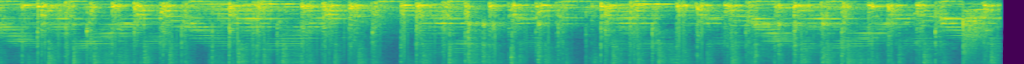}
\\source: this instrumental song features a [relaxing] melody with a country feel accompanied by a guitar piano simple percussion and bass in a slow tempo \\
\includegraphics[width=0.8\textwidth, height=0.1\textheight]{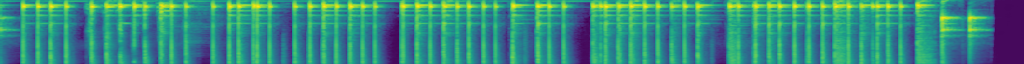} 
\\edit: this instrumental song features a [cheerful] melody with a country feel accompanied by a guitar piano simple percussion and bass in a slow tempo\\
\caption{Editing Type 8 Extract Instrument}
\label{fig:type8}
\end{figure*}

\section{safeguards}
\label{H}
In the processing of the data and models involved in this study, we fully considered the potential risks. We ensure that all data sources are rigorously screened and vetted, and the model we used is absolutely trained from the safe dataset to minimize the security risks of being misused.
\end{document}